\documentclass[aip,apl,reprint]{revtex4}
\usepackage{amssymb,amsmath}
\usepackage{pgf}
\usepackage{color}
\usepackage{subfigure}
\usepackage{epstopdf}
\usepackage{graphicx}% Include figure files
\usepackage{dcolumn}% Align table columns on decimal point
\usepackage{bm}% bold math
\usepackage[mathlines]{lineno}% Enable numbering of text and display math
\usepackage{hyperref}

\begin{document}

\title[]{Far-field super-resolution imaging with a planar hyperbolic metamaterial lens beyond the Fabry-Perot resonance condition}
%\shorttitle{Title} %Insert here a short version of the title if it exceeds 70 characters

\author{Cheng Lv}
\affiliation{Key Laboratory of Terahertz Solid-State Technology, Shanghai Institute of Microsystem and Information Technology, Chinese Academy of Sciences, Shanghai 200050, China}
  \affiliation{State Key Laboratory of Functional Materials for Informatics, Shanghai Institute
of Microsystem and Information Technology, Chinese Academy of Sciences, Shanghai 200050, China}
\affiliation {University of Chinese Academy of Sciences, Beijing 100049, People's Republic of China}
\author{ Wei Li}
\email{waylee@mail.sim.ac.cn}
  \affiliation{State Key Laboratory of Functional Materials for Informatics, Shanghai Institute
of Microsystem and Information Technology, Chinese Academy of Sciences, Shanghai 200050, China}

\author{Xunya Jiang}
\affiliation{Dept. of Illuminating Engineering and Light Sources, School of Information Science and Engineering, Fudan University, Shanghai 200433, China}
\affiliation{Engineering Research Center of Advanced Lighting Technology, Fudan University, Ministry of Education, Shanghai 200433, China}

\author{Juncheng Cao}
\email{jccao@mail.sim.ac.cn}
\affiliation{Key Laboratory of Terahertz Solid-State Technology, Shanghai Institute of Microsystem and Information Technology, Chinese Academy of Sciences, Shanghai 200050, China}

\begin{abstract}
 We demonstrate achieving the far-field super-resolution imaging can be realized by using a planar hyperbolic metamaterial lens (PHML), beyond the Fabry-Perot resonance condition. Although the thickness of the PHML is much larger than wavelength, the PHML not only can transmit radiative waves and evanescent waves with high transmission, but also can collect all the waves in the image region with the amplitudes of them being the same order of magnitude. We present a design for a PHML to realize the far-field super-resolution imaging, with the distance between the sources and the images 10 times larger than wavelength. We show the superresolution of our PHML is robust against losses, and the PHML can be fabricated by periodic stacking of metal and dielectric layers.
 \end{abstract}

\maketitle

\section{Introduction}
Remarkable progress has been made over the past decade in achieving optical super-resolution imaging\cite{2012NatMaterials} that breaks the traditional resolution limit,
offering substantial advances for the imaging and lithography systems that are the corner stones of modern biology and electronics.
Metamaterial-based superlenses have shown the super-resolution imaging in the near field\cite{2000PRL85Pendry,2005Science308NicholasFang},
but magnification of subwavelength features into the far field has not been possible.
Recently, super-oscillatory lenses\cite{2009NANOLETTERS9Nikolay} and hyperlenses\cite{2007Science315ZhaoweiLiu,2007Science315Smolyaninov}
have demonstrated to the far-field super-resolution imaging. However, both the super-oscillatory lenses and hyperlenses suffer from some limitations:
that super-oscillatory lenses come at a price of losing most of the optical energy into diffuse sidebands;
and hyperlenses need complex co-centrically curved-layer structures, which not only bring more challenges in fabrication, but also fundamentally lack the Fourier transform function due to their inability to focus plane waves\cite{2012NatureCommunications3DylanLu}. To avoid these limitations, planar hyperbolic metamaterial lenses (PHMLs) were reported\cite{2006PRB73PavelABelov,2008OE16ChangtaoWang}
and have demonstrated capable of providing not only Fourier transform function
by introducing a phase compensation mechanism\cite{2012NatureCommunications3DylanLu},
but also advantages of super-resolving capabilities. However, so far most of the studies for the super-resolution imaging of PHMLs are done in the near field
\cite{2006PRB73PavelABelov,2008OE16ChangtaoWang}
(with the distances between source and image are much smaller than unit wavelength). Recently, P. Belov and co-authors proposed an alternative regime based on the Fabry-Perot resonance effect, and thus it is capable of transmitting images to far-field \cite{2008PRB77BelovParini}. Nevertheless, the regime itself assumes that the thickness of the PHML should be equal to an integer number of half-wavelengths \cite{2008PRB77BelovParini}. In this paper, we demonstrate that the far-field super-resolution imaging also arises in well-designed PHMLs when the Fabry-Perot resonance condition does not hold.

Generally, achieving the far-field super-resolution imaging through a planar lens is very challenging.
For the subwavelength imaging, both evanescent waves and radiative waves are required,
and their amplitudes should be comparable with each other in the image.
If the amplitude of radiative waves is much larger than that of evanescent waves,
the subwavelength information taken by evanescent waves would be concealed by radiative waves,
as thus the imaging system loses the super-resolution capability\cite{InContrast}.
Therefore, the amplitude of radiative waves and evanescent waves should be the same order of magnitude in the image region for superresolution.
But satisfying this condition is generally very difficult for the far-field imaging,
because of the inherent nature of evanescent waves with amplitudes exponential decaying in the air\cite{1986Electromagneticwavetheory,2011APL100WeiLi}.
One of the most striking properties of hyperbolic metamaterials is their strong ability to carry evanescent waves to far field \cite{2007Science315ZhaoweiLiu},
which enables PHMLs to be good candidates for far-field super-resolution imaging.
To realize the far-field superresolution, however, a PHML should overcome at least four difficulties:
First, beyond the Fabry-Perot condition, the PHML should be able to transmit radiative waves and evanescent waves from a source with high transmission and low loss.
Second, it should be able to collect the radiative waves and evanescent waves in the image, with the amplitudes of them being the same order of magnitude.
Third, the distance between the source and image should be much larger than unit wavelength.
Fourth, the resolution of the PHML must be robust against losses to a certain degree, including radiative losses and material losses.

In this work, we present our design for a PHML that can be used to realize the far-field super-resolution imaging beyond the Fabry-Perot condition, with the resolution less than $0.1$ wavelength and simultaneously the distance between the image and the source as far as 10 wavelength or even farther. We show our PHML can be made from subwavelength layered structure, which is robust against losses. For simplicity, we constrain ourselves to only study the PHML's far-field resolution properties in this paper.

\section{Model}
\begin{figure}%[H]
\centering
\includegraphics[width=0.75\columnwidth]{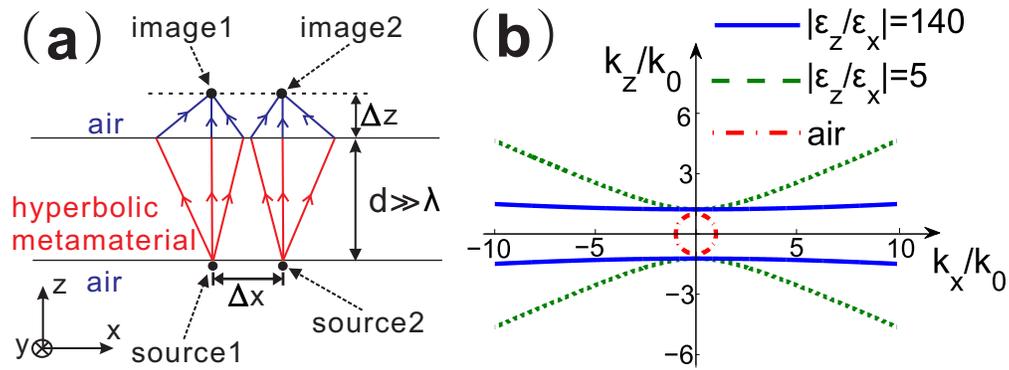}
%\onefigure{Fig1.eps}
\caption{\label{fig.1} (a)The schematic diagram of our model.
(b) The EFCs of hyperbolic metamaterials with $|\varepsilon_z/\varepsilon_x|=140$ (blue solid line) and $|\varepsilon_z/\varepsilon_x|=5$ (green dash line), and the EFC of air (red dash-dot line).}
\end{figure}

Our model is shown in Fig.\ref{fig.1}(a), in which a slab-like planar hyperbolic metamaterial with thickness $d$ ($d$ is much larger than the wavelength, but doesn't hold the Fabry-Perot condition) is positioned at $0<z<d$ for imaging,
two monochromatic line sources with infinity length along $y$ direction are set very close to the lower interface and their images are formed above the upper interface at a distance of $\Delta z$.
The distance between the two sources is $\Delta x$. The hyperbolic metamaterial has a matrix form permittivity
$\varepsilon=diag(\varepsilon_x,\varepsilon_y,\varepsilon_z)$ with $\varepsilon_x=\varepsilon_y$,
and $\varepsilon_x \cdot \varepsilon_z<0$,
and so its dispersion relation $k_{x}^2/\varepsilon_{z}+k_{z}^2/\varepsilon_{x}=k_0^2$ is hyperbolic \cite{2008OE16ChangtaoWang, 2012APL100WeiLi},
where $k_0=2\pi/\lambda$ is the wavevector, with $\lambda$ the wavelength in air.
The equi-frequency contours (EFCs) of hyperbolic metamaterials are hyperbolic. In the case of $\varepsilon_z<0$ and $\varepsilon_x>0$, two typical examples are shown in Fig.\ref{fig.1}(b). In this work, only the transverse electric (TE) mode (i.e., $H=H_y\neq 0$) is considered.

\begin{figure}%[H]
\centering
\includegraphics[width=0.75\columnwidth]{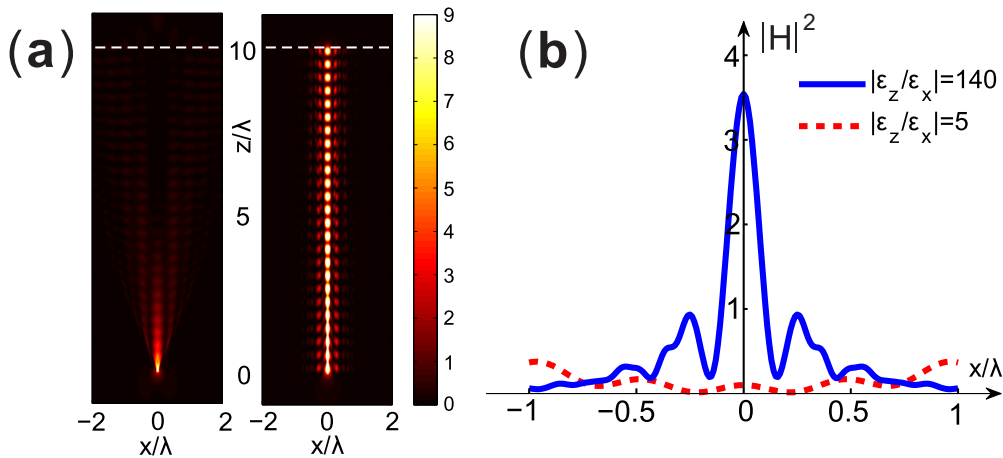}
\caption{\label{fig.2}
(a) The distribution of $|\vec{H}|^2$ of a single source placed at the interface of PHMLs with $|\varepsilon_z/\varepsilon_x|=5$ (left panel), and $|\varepsilon_z/\varepsilon_x|=140$ (right panel). Both lenses' width are $d=10\lambda$.
(b) The distribution of $|\vec{H}|^2$ at the interface (along the white dash line in Fig2.(a)) for the PHMLs
with $|\varepsilon_z/\varepsilon_x|=140$ (blue solid line) and $|\varepsilon_z/\varepsilon_x|=5$ (red dash line), respectively.}
\end{figure}

\section{Choice of hyperbolic metamaterials}
For a PHML, there are many choices of the transverse axis' direction of the hyperbolic EFC. In this work we choose the one that is parallel to the $z$ axis (corresponding to $\varepsilon_z<0$ and $\varepsilon_x>0$ as shown in Fig.1(b)), we find it is much suitable for the far-field imaging.
If the transverse axis is perpendicular to the $z$ axis (in this case we have $\varepsilon_z>0$ and $\varepsilon_x<0$), the radiative waves in air with wavevector $k_x<\sqrt{\varepsilon_z}k_0$ would exponentially decay in the lens, so this case is unsuitable for far-field imaging. If the transverse axis is tilted of an angle with $z$ axis\cite{2012PRB86GiuseppeCastaldi}, this kind of PHML can be directly used as a Fourier transform lens, but it's also unsuitable for far-field superresolution because of its large lateral deviation of the image through a long distance.

After fixing the transverse axis, i.e., $\varepsilon_z<0$ and $\varepsilon_x>0$, the value $|\varepsilon_z/\varepsilon_x|$ determining the eccentricity of the hyperbola is the core parameter of our system. We choose large $|\varepsilon_z/\varepsilon_x|$ for our PHML. There are two reasons for our choice. First, if $|\varepsilon_z/\varepsilon_x|$ is not large enough, the image will be formed at a long distance $\Delta z$ from the interface of the lens\cite{2009PRB79AFang}. In this case the evanescent waves would be too small to carry subwavelength information for the long-distance imaging. Second, for a PHML, larger $|\varepsilon_z/\varepsilon_x|$ corresponds to better self-collimation of light, which is very significant for the super-resolution imaging. Hyperbolic metamaterials with larger $|\varepsilon_z/\varepsilon_x|$ have flatter EFCs, and so the light diffraction can be lower when lights propagate in the hyperbolic metamaterials with larger $|\varepsilon_z/\varepsilon_x|$. To show this, we present two typical examples, i.e., one case is $|\varepsilon_z/\varepsilon_x|=5$ (with $\varepsilon_x=1.5 ,\varepsilon_z=-7.5$), and another case is $|\varepsilon_z/\varepsilon_x|=140$ (with $\varepsilon_x=1.5, \varepsilon_z=-210$), whose EFCs are shown in Fig.\ref{fig.1}(b). From this figure, we can see the latter one has much flatter EFC than that of the former one. As a comparison, the two hyperbolic metamaterials are used for imaging, according to the model shown in Fig.\ref{fig.1}(a), with $\Delta x=0$, $d=10\lambda$.
Their magnetic field intensity distributions calculated by Green's function method\cite{1986Electromagneticwavetheory,2012JKPS60WLi,2013PRA87ZLiu} are shown in Fig.2, in which Fig.\ref{fig.2}(a) shows the distributions in the $x$-$z$ plane, and Fig.\ref{fig.2}(b) shows the distributions in an imaging region along the upper interface of the hyperbolic metamaterials. Comparing the left panel with the right panel in Fig.\ref{fig.2}(a), we can see the right one (corresponding to the latter one with $|\varepsilon_z/\varepsilon_x|=140$) has much better capable of self-collimation. Also, from Fig.\ref{fig.2}(b), we can see the intensity of the latter one's image is more than 35 times larger than that of the former one, which indicates the latter one can focus more waves in the image. Therefore, hyperbolic metamaterials with larger $|\varepsilon_z/\varepsilon_x|$ are more suitable for super-resolution imaging.

\section{Far-field superresolution}
\begin{figure}%[H]
\centering
\includegraphics[width=0.75\columnwidth]{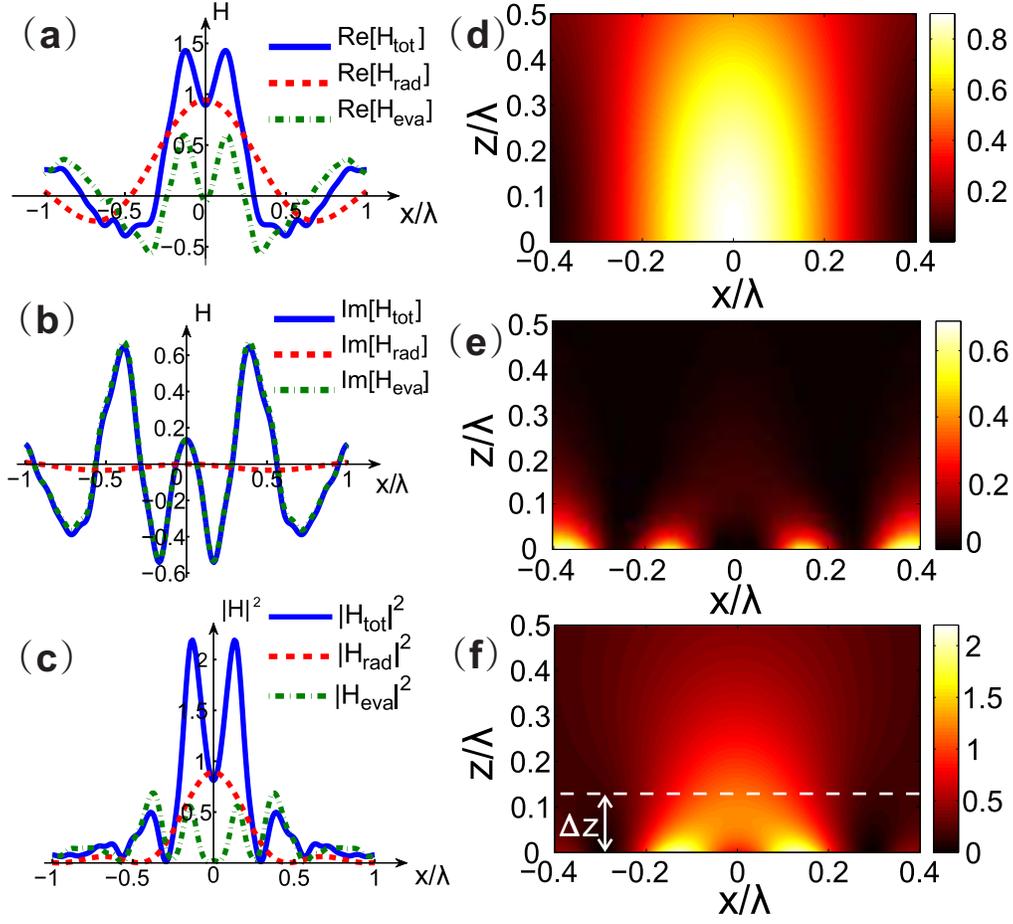}
\caption{\label{fig.3} (a)-(c) The field distributions along the upper interface of our PHML. (a) The real part of $\vec{H}_{tot}$ (blue solid line), $\vec{H}_{rad}$ (red dash line) and $\vec{H}_{eva}$ (green dash-dot line). (b) The imaginary part of $\vec{H}_{tot}]$ (blue solid line), $\vec{H}_{rad}$ (red dash line) and $\vec{H}_{eva}$ (green dash-dot line). (c) The distribution of $|\vec{H}_{tot}|^2$ (blue solid line), $|\vec{H}_{rad}|^2$ (red dash line) and $|\vec{H}_{eva}|^2$ (green dash-dot line).
(d)-(f) are the distribution of $|\vec{H}_{rad}|^2$, $|\vec{H}_{eva}|^2$ and $|\vec{H}_{tot}|^2$ in the image region, respectively. In (a)-(f), $\gamma=0.01$.}
\end{figure}

In Fig.\ref{fig.2}(b), the full width half maximum (FWHM) of the intensity distribution for the case of $|\varepsilon_z/\varepsilon_x|=140$ (blue solid line) is about $0.2\lambda$ that is much smaller than half of wavelength. In addition, the distance between the source and the image are $10\lambda$ that is much larger than unit wavelength, suggesting our system has the ability for the far-field subwavelength imaging.

To show it more clearly, we use such a PHML for imaging. As shown in Fig.\ref{fig.1}(a), the two sources are spaced at a distance of $\Delta x=0.3\lambda$, the thickness of the PHML is $d=10\lambda$. For the hyperbolic metamaterial, the real part of the relative permittivity is $\emph{Re}[\varepsilon_x]=1.5$ and $\emph{Re}[\varepsilon_z]=-210$ (i.e., $\emph{Re}[\varepsilon_z]/\emph{Re}[\varepsilon_x]=140$), and the imaginary part is much smaller than the real part, which is phenomenologically introduced by
\begin{equation}
\emph{Im}[\varepsilon_{x(z)}]=\gamma|\emph{Re}[\varepsilon_{x(z)}]|,
\end{equation}
where $\gamma$ is the loss coefficient. Based on the Green's function method, we calculated the field ($H_{rad}$, $H_{eva}$ and $H_{tot}$) distributions and the intensity distributions in the image region, as shown in Fig.\ref{fig.3}(a)-(e). Here $H_{rad}$, $H_{eva}$ and $H_{tot}=H_{rad}+H_{eva}$ represent the radiative waves, the evanescent waves, and the total field, respectively. Note that $H_{rad}$, $H_{eva}$ and $H_{tot}$ are all complex. Figure 3(c) shows our PHML's ability for the far-field super-resolution imaging. In this figure, we show the intensity distributions of $|H_{rad}|^2$, $|H_{eva}|^2$, and $|H_{tot}|^2$ in the image region that is very close to the upper interface of the PHML ($\Delta z\sim 0$). In Fig.\ref{fig.3}(c), for the intensity distribution of the total field (corresponding to the blue solid line), it's very clear to see that the two images (corresponding to the two main peaks) can be well resolved, although their distance $\Delta x=0.3\lambda$ is smaller than $0.5\lambda$. Furthermore, we stress that the distance between the sources and the images is $d+\Delta z\sim 10 \lambda$, which is much larger than $\lambda$. So our PHML does have the ability for the far-field super-resolution imaging.

Physically, the reason that our PHML exhibits the far-field super-resolution capability can be explained as follows. First, as shown in Fig.\ref{fig.3}(a) and Fig.\ref{fig.3}(b), both radiative waves and evanescent waves from sources can be transmitted through our PHML with high transmission (although the thickness of the PHML is very large), and they can be collected at the image region. Second, also from Fig.\ref{fig.3}(a)-(c), we can see the amplitudes of the radiative waves and the evanescent waves are comparable with each other, and they are constructive at the image peaks but simultaneously destructive at the side peaks.

\begin{figure}%[H]
\centering
\includegraphics[width=0.75\columnwidth]{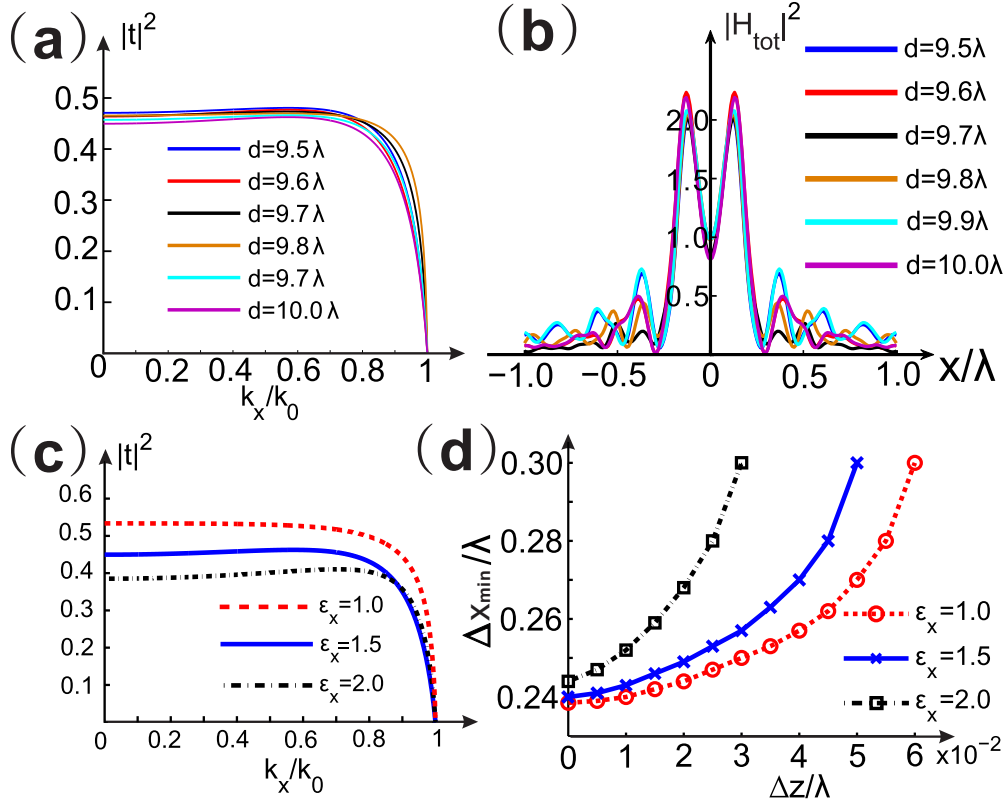}
\caption{\label{fig.4} (a) The transmission rate of the PHML with different thicknesses. (b) The $|H_{tot}|^2$ distribution at the interface of the PHMLs with different thicknesses. Other parameters in (a) and (b) are the same as Fig.3. (c) The transmission of the PHML with $\varepsilon_x=1.0$ (red dash line), $\varepsilon_x=1.5$ (blue solid line) and $\varepsilon_x=2.0$ (black dash-dot line). (d) $\Delta x_{min}$ vs $\Delta z$ with $\varepsilon_x=1.0$ (red dash line), $\varepsilon_x=1.5$ (blue solid line) and $\varepsilon_x=2.0$ (black dash-dot line). In (c) and (d), the thickness of the PHML $d=10\lambda$, and eccentricity $|\varepsilon_z/\varepsilon_x|=140$ are fixed.}
\end{figure}

Evanescent waves play a very important role on the super-resolution. Without evanescent waves, our PHML would lose its super-resolution capability. In Fig.\ref{fig.3}(c), the images formed only by the radiative waves become a large spot without any subwavelength information that totally cannot be resolved. This fact can also be found in Fig.\ref{fig.3}(d)-(f). In Fig.\ref{fig.3}(d), the images formed only by the radiative waves cannot be resolved at all in the image region; while in Fig.\ref{fig.3}(f), the two images can be resolved well in the image region within a certain range of $\Delta z$ with the help of the evanescent waves. As $\Delta z$ increasing, the amplitudes of evanescent waves would exponentially decay that lessens the resolution of the PHML. As shown in Fig.\ref{fig.3}(f), when $\Delta z>0.05\lambda$, the two images cannot be resolved at all. To show the increase of $\Delta z$ can impair the resolution of the PHML (i.e., the minimal distinguishable distance $\Delta x_{min}$ would become larger as $\Delta z$ increases), $\Delta x_{min}$ versus $\Delta z$ are plotted in Fig.4(d) (blue line). When $\Delta z=0$, our PHML can achieve its highest resolution, with the minimal distinguishable distance $\Delta x_{min}=0.24\lambda$. In Fig.\ref{fig.4}(d), whether the images can be resolved or not is following the Rayleigh criterion\cite{1964PrinciplesofOptics}. In this work, we choose the criterion of distinguishable two images should be smaller than the critical contrast $I_{sad}/I_{max}=75\%$, where $I_{sad}$ and $I_{max}$ represent the saddle point intensity and maximum intensity, respectively. Our criterion is a little bit stricter than that used in ref\cite{2009NANOLETTERS9Nikolay}.

\section{Far-field superresolution beyond the Fabry-Perot resonance condition}
Our PHML can be capable of far-field superresolution beyond the Fabry-Perot resonance condition. To see it, here we would like to show the far-field superresolution ability of several our PHMLs with different thicknesses (i.e., $d=9.5$, $9.6$, $...$, $10.0\lambda$) as shown in Fig.\ref{fig.4}, in which the Fabry-Perot resonance condition is satisfied only when $d=9.8\lambda$, while other thickness values correspond to the nonresonant case. From Fig.\ref{fig.4}(b), by comparing with the $|H_{tot}|^2$ distributions in the resonant case and those in the nonresonant case, we can see their $|H_{tot}|^2$ distributions are very similar and all the PHMLs are able to superresolution. Also, from Fig.\ref{fig.4}(a), we can see that the transmissions (here we only plot the radiative waves) of the PHMLs with different thicknesses are similar, too. These similarities of superresolution and transmission between the PHMLs with different thicknesses indicate that the far-field super-resolution ability of our PHMLs is not linked to the Fabry-Perot resonance effect, but related to the nonresonant impedance matching.

We find in our case, better (nonresonant) impedance matching corresponds to better superresolution, as shown in Fig.\ref{fig.4}(c) and (d). For the PHMLs in Fig.\ref{fig.4}(c) and (d), the eccentricity $|\varepsilon_z/\varepsilon_x|=140$ is fixed, but the values of $\varepsilon_x$ are different, i.e., with $\varepsilon_x=1.0$, $\varepsilon_x=1.5$, and $\varepsilon_x=2.0$, respectively. Other parameters are the same as Fig.\ref{fig.3}. Clearly, when $\varepsilon_x$ becomes closer to unit, the nonresonant impedance matching will become better. From Fig.\ref{fig.4}(d) we can find when the impedance is matching better, the minimal distinguishable distance $\Delta x_{min}$ becomes smaller, and the largest $\Delta z$ becomes larger, which corresponds to better superresolution. In the following, we will show that to realize a PHML with quasi-perfect nonresonant impedance matching is not very difficult.

\section{Robust against losses}
\begin{figure}%[H]
\centering
\includegraphics[width=0.75\columnwidth]{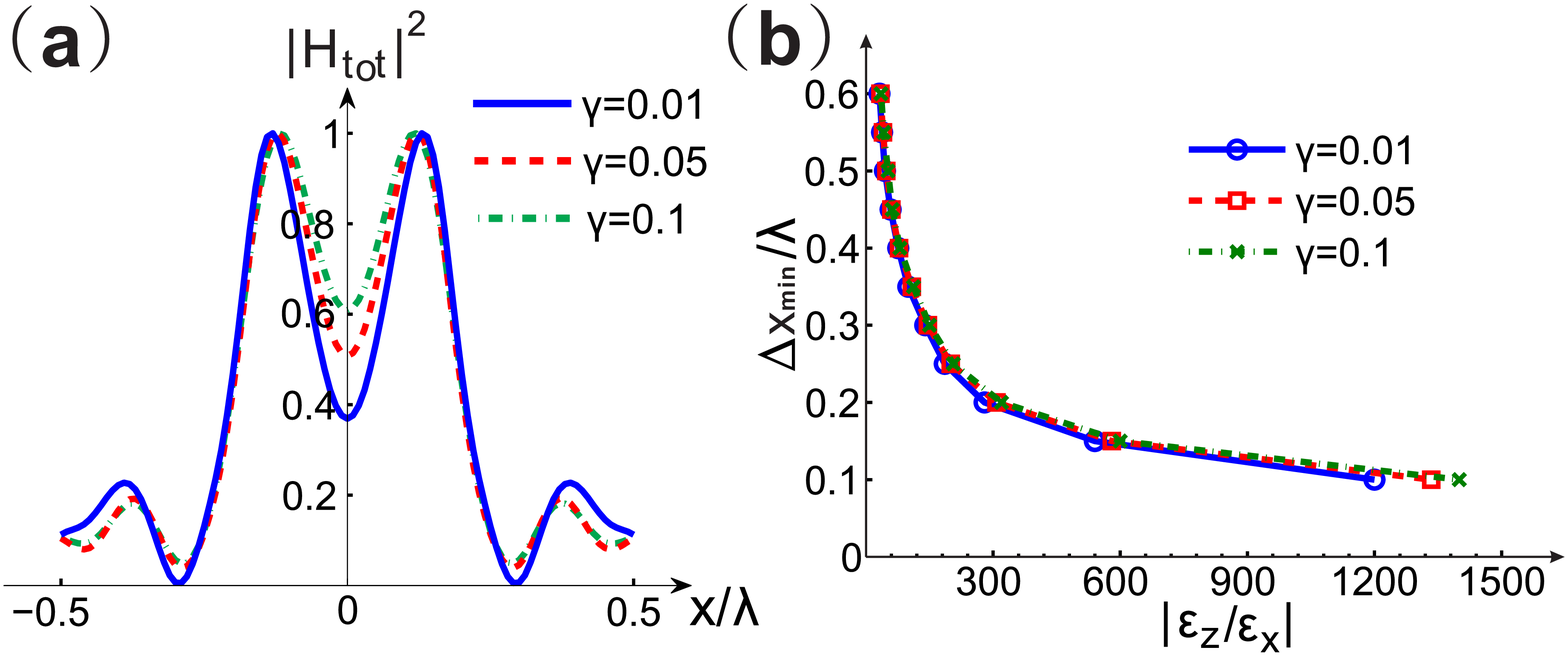}
\caption{\label{fig.5} (a) The normalized intensity distribution at $z=10\lambda$ with $\gamma=0.01$ (blue solid line), $\gamma=0.05$ (red dash line) and $\gamma=0.1$ (green dash-dot line). Other parameters are the same as Fig.3.
(b) The minimal distinguishable distance $\Delta x_{min}$ vs $|\varepsilon_z/\varepsilon_x|$ with
$\gamma=0.01$ (blue solid line), $\gamma=0.05$ (red dash line) and $\gamma=0.1$ (green dash-dot line). }
\end{figure}

The superresolution of our PHML has robustness against losses, which is also important for imaging system. To see it, three PHMLs with three different losses are respectively used for imaging. The losses are phenomenologically introduced by Eq.(1), with the loss coefficient $\gamma=0.01$, $0.05$ and $0.1$, respectively. The intensity distributions of the three PHMLs' image region at $\Delta z=0$ are shown in Fig.\ref{fig.5}(a). In Fig.\ref{fig.5}(a), the parameters are the same as Fig.3, except the loss coefficient $\gamma$. In this figure, the maximum intensity value of each intensity distribution has been normalized to unit. From Fig.\ref{fig.5}(a), we can see even for a big loss ($\gamma=0.1$), our PHML also has the far-field super-resolution capability. For our PHMLs with different $|\varepsilon_z/\varepsilon_x|$, the robustness against losses are also valid, as shown in Fig.\ref{fig.5}(b). From Fig.\ref{fig.5}(b), we can also see that the larger $|\varepsilon_z/\varepsilon_x|$, the higher resolution of our PHML, which agrees with our previous discussion.

The fact that the superresolution ability of the PHML is robust against losses doesn't mean the losses has nothing to do with the transmitted light waves. Actually, the losses play a very important role in the amplitude of the transmitted light waves. For such a thick PHML (e.g. $d\sim 10\lambda$), the intensities of the transmitted light waves with different losses can differ by orders of magnitude. For instance, in Fig.5(a), the light's intensity magnitude in the case with $\gamma=0.01$ can be $\sim$$10^2$ times larger than that with $\gamma=0.1$. Here we would like to emphasize that, although the losses can lead to large attenuation of signal strength, the losses don't impair the superresolution. This result is meaningful. According to it, when using some techniques such as loss-compensation to overcome the losses, we can predict that the PHML will retain its superresolution ability.

\section{Anisotropic layered structure for PHMLs}
\begin{figure}%[H]
\centering
\includegraphics[width=0.75\columnwidth]{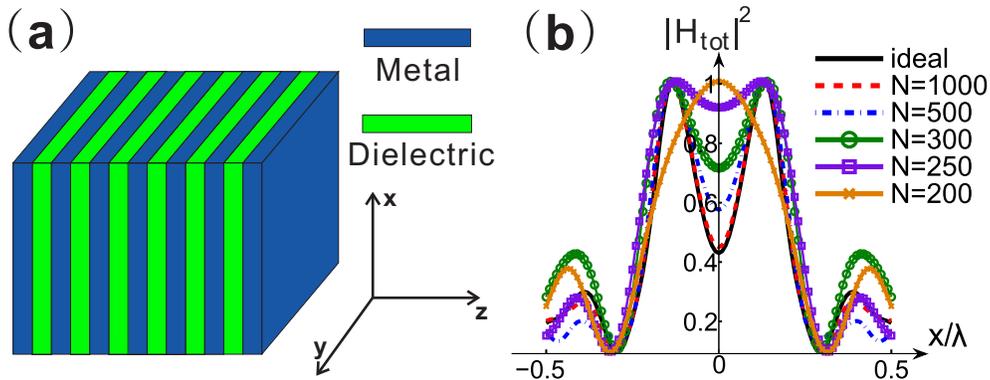}
\caption{\label{fig.6} (a) The schematic diagram of the effective hyperbolic lens consisting of alternately layered metal and dielectric layers with total thickness $d$.
 (b) The normalized distribution of $|\vec{H}_{tot}|^2$ at $z=10\lambda$ with $N=200,250,300,500,1000$
and distribution calculated with ideal hyperbolic media (black solid line).}
\end{figure}

Our PHMLs can be realized by anisotropic structure comprising alternately layered metal and dielectric films, as shows in Fig.\ref{fig.6}(a). The thickness of the cell (a cell means a combination of one metal layer and one dielectric layer) is much smaller than the wavelength. Using the transfer-matrix method and imposing the Bloch theorem\cite{2012APL100WeiLi}, the effective permittivity of such a layered structure can be obtained as\cite{2006PRB73PavelABelov} $\varepsilon_x=\varepsilon_mf_m+\varepsilon_d f_d$ and $\varepsilon_z^{-1}=f_m/\varepsilon_m+f_d/\varepsilon_d$. Here $\varepsilon_m$ and $\varepsilon_d$ are the permittivities of the metal and dielectric, respectively. $f_m$ and $f_d=1-f_m$ are the filling factors for metal and dielectric layer, respectively. In realization, the number of cells of the layered structure is finite. For a practical PHML, the number of cells has much influence on the super-resolution imaging. To see it, here we present a comparison of the $|\vec{H}_{tot}|^2$ distributions at the upper interface of an ideal PHML with $\varepsilon_x=1.5$, $\varepsilon_x=-210$ and the practical PHMLs with different cell number $N$ in Fig.\ref{fig.6}(b). All the PHMLs in Fig.\ref{fig.6}(b) have a fixed thicknesses $d=10\lambda$. For the practical PHMLs with different cell number $N$, the thicknesses of each cell $d/N$ are different, but in each cell the filling factors $f_d=0.52$, $f_m=0.48$ and the permittivities $\varepsilon_d=12.25$, $\varepsilon_m=-10.16+0.03i$ are fixed. In order to show the resolution ability of each PHML, the maximum intensity value of each intensity distribution in Fig.\ref{fig.6}(b) is also normalized to unit. From Fig.\ref{fig.6}(b), we can see that the superresolution can be reached with the cell number $N=300$, and when N increases, the distribution of $|\vec{H}_{tot}|^2$ of the layered structure becomes more similar to that of the ideal one, which is expected.

Next, we will discuss the feasibility of the practical PHML achieves higher resolution. According to Fig.\ref{fig.5}(b), we can see that the resolution can be $0.1\lambda$ or even smaller when $|\varepsilon_z/\varepsilon_x|>1200$. This can be realized by choosing the suitable parameters ($f_m,f_d,\varepsilon_m,\varepsilon_d$) for the multilayered structure as shown in Fig.\ref{fig.6}(a). For example, for the 1000-cell multilayered structure shown in Fig.\ref{fig.6}(b), with the parameters remaining unchanged except the filling factors, i.e., they become $f_m=0.545$ and $f_d=1-f_m=0.455$, then the effective permittivities $\varepsilon_x$ and $\varepsilon_z$ of the 1000-cell multilayered structure become about 0.037 and -60.31, respectively, and the eccentricity $|\varepsilon_z/\varepsilon_x|$ is about 1638. Using such 1000-cell multilayered structure as the PHML, we find the resolution is about $0.098\lambda$.

\section{Conclusion}
In conclusion, we present an approach to achieve superresolution at a long-distance ($10\lambda$) through a PHML beyond the Fabry-Perot resonance condition, which can distinguish two sources with distance $0.1\lambda$. The resolution of our imaging system is robust against losses. The PHMLs can be structured by periodic stacking of metal and dielectric layers.

\emph{\textbf{Acknowledgments}}
This work was supported by the National Natural Science Foundation of China (Grant Nos. 11204340, 61131006, 61021064, and 11174309), and the Shanghai Municipal Commission of Science and Technology (Project Nos. 11ZR1443800, and 10JC1417000).


\begin{thebibliography}{0}
\bibitem{InContrast}In contrast, if the amplitude of evanescent waves is much larger than that of radiative waves, due to the large evanescent waves, the image region can have obvious subwavelength feature, but simultaneously the corresponding intensity distribution have many side peaks in the image region,
whose amplitudes are as large as the main peak, severely impairing the imaging.
\end{thebibliography}
\end{document}